\documentclass[12pt]{iopart}

\usepackage{epstopdf}

\usepackage{cite}

\usepackage[dvipdfmx]{graphicx}

\newcommand{\bib}{\bibitem}
\newcommand{\todayd}{\the\year/\the\month/\the\day}
\newcommand{\eq}[1]{\begin{equation} #1 \end{equation}}
\newcommand{\eqa}[2]{\begin{equation} #1 \label{#2} \end{equation}}
\newcommand{\del}{\partial}

\newcommand{\la}{\left\langle}
\newcommand{\ra}{\right\rangle}

\newcommand{\balign}[1]{\begin{eqnarray} #1 \end{eqnarray}}

\newcommand{\lb}{\label}
\newcommand{\nt}{\nonumber}

\newcommand{\eqref}[1]{(\ref{#1})}
\newcommand{\eqqref}[1]{Eq.~\eqref{#1}}
\newcommand{\ffref}[1]{Fig.~\ref{#1}}

\newcommand{\figin}[4]
{
\begin{figure}[tb]\centering\includegraphics[width= #1]{#2}\caption{#3}\label{#4}\end{figure}
}

\def \({\left(}
\def \){\right)}

\def\rnum#1{\resizebox{0.5em}{\height}{\expandafter{\romannumeral #1}}}
\def\Rnum#1{\resizebox{0.5em}{\height}{\uppercase\expandafter{\romannumeral #1}}}

\makeatletter
  \newcommand{\subsubsubsection}{\@startsection{paragraph}{4}{\z@}%
    {1.0\Cvs \@plus.5\Cdp \@minus.2\Cdp}%
    {.1\Cvs \@plus.3\Cdp}%
    {\reset@font\sffamily\normalsize}
  }
  \makeatother
\makeatletter
\def\verbatim@font{\normalfont\fontfamily{txr}\selectfont
\let\do\do@noligs
\verbatim@nolig@list}
\makeatother

\begin{document}

\title{Role of measurement-feedback separation in autonomous Maxwell's demons}
\author{Naoto Shiraishi$^1$, Sosuke Ito$^2$, Kyogo Kawaguchi$^2$, and Takahiro Sagawa$^1$}
\address{$^1$ Department of basic science, The University of Tokyo, 3-8-1 Komaba, Meguro-ku, Tokyo, Japan}
\address{$^2$ Department of Physics, The University of Tokyo, 7-3-1 Hongo, Bunkyo-ku, Tokyo, Japan}
\ead{shiraishi@noneq.c.u-tokyo.ac.jp}
\vspace{10pt}
\begin{indented}
\item \todayd
\end{indented}

\begin{abstract}
We introduce an information heat engine that is autonomous (i.e., without any time-dependent parameter) but has separated measurement and feedback processes. 
This model serves as a bridge between different types of information heat engines inspired by Maxwell's demon; from the original Szilard-engine type systems to the autonomous demonic setups. 
By analyzing our model on the basis of a general framework introduced in our previous paper [N. Shiraishi and T. Sagawa, Phys. Rev. E {\bf 91}, 012130 (2015).], we clarify the role of the separation of measurement and feedback in the integral fluctuation theorems.
\end{abstract}

\section{Introduction}\lb{s:intro}

``Maxwell's demon'' is a thought experiment proposed by J. C. Maxwell~\cite{Maxwell}: if a thermodynamic system is subjected to feedback control at the level of thermal fluctuations, the second law can be apparently violated. 
A prominent example is the Szilard engine~\cite{Szilard}, which is a composite system of an engine and a memory. 
The memory measures the state of the engine and performs feedback to the engine. 
The feedback procedure allows positive work to be extracted from the engine through an isothermal cyclic process, which, under the usual circumstance, is prohibited by the second law of thermodynamics. 
Szilard characterized the two separated steps of measurement and feedback as the processes which change the extent of the correlation between the engine and the memory, and suggested the possibility of an extended framework of thermodynamics for systems with the change in correlation.
The consistency of the second law of thermodynamics with the existence of Maxwell's demon has been discussed vigorously~\cite{discuss}.
It has been suggested that the key to understand the consistency is the change in the volume of the phase space~\cite{Magnasco}, which characterizes the thermodynamic irreversibility~\cite{KPB}.
Modern theories have revealed that this property is captured by the mutual information, which is a quantity that measures the correlation between the engine and the memory.
The generalized thermodynamic relations with the mutual information has been discussed for a single feedback process~\cite{Touchette, SU2010}, continuous feedback processes~\cite{Cao, HV, SU2010full}, feedback cooling~\cite{Kim, ItoSano, Martin}, and more general information processing~\cite{Granger, SU2012,SU2012full}.

Although these information-theoretic frameworks are expected to be useful in various problems in small fluctuating systems such as biochemical sensing~\cite{Tostevin, Lan} and quantum mesoscopic systems~\cite{Esposito1, e-box}, there is a critical remark to be made; the original framework, which is applicable to the Szilard-type demons with step-by-step separated measurement and feedback processes, cannot be directly applied to autonomous stochastic information processes. 
This is because the processes of measurement and feedback can themselves occur stochastically at any time, and thus they are inseparable in autonomous setups. 
In purpose to extend the applicability of information thermodynamics to autonomous Maxwell's demons, intensive researches on the model of autonomous demons~\cite{Sekimoto, MJ, HSP, Esposito1} and information flow~\cite{Marko, Schreiber, Liang, Majda, Allahverdyan, IS, BS2013, Seifert2014, Jordan2014, JS, SS} have been recently made. 
Autonomous demons can be modeled as bipartite Markovian systems where the measurement and the feedback processes are not, in general, separated. 
The integral fluctuation theorem (IFT) obtained for the autonomous demons has distinct features from the relation obtained for the Szilard-type systems~\cite{SU2012, SU2012full}. 
The crucial difference comes from the fact that in the Szilard-type systems, the state of the engine (memory) is fixed externally during the measurement (feedback) phase, whereas in autonomous demons, it is not.

In this paper, we clarify the difference between Szilard-type demons and autonomous demons on the basis of a general framework introduced in our previous paper~\cite{SS}. 
In purpose to build a bridge between these two different setups, we propose a new model of Maxwell's demon, which is autonomous but possesses separated measurement and feedback processes. 
The key ingredient in the model is an additional stochastic variable, which plays a role of separating the measurement phase and the feedback phase. 
An IFT with the mutual information can be derived for this model, which is more similar to another IFT satisfied in Szilard-engine type demons~\cite{SU2012, SU2012full} than to that for autonomous bipartite demons~\cite{SS}.

This paper is organized as follows. 
In Sec.~\ref{s:problem}, we review IFTs for Szilard-type demons and autonomous demons, where the mutual information plays a crucial role. 
We also discuss a general framework that leads to IFT of autonomous and non-autonomous demons in a unified way, on the basis of the concept of the partial entropy production.
In Sec.~\ref{s:class1}, we introduce a model of the autonomous demon with separated measurement and feedback phases, and derive an IFT for this model. 
On the basis of these discussions, in Sec.~\ref{s:class1-class2}, we elucidate the role played by the separation of the measurement phase and the feedback phase. 

\section{Information thermodynamics: a brief review}\lb{s:problem}

\subsection{Thermodynamics of small systems}\lb{s:gen-set}

Throughout this paper, we consider Markov jump processes during the time interval $0\leq t\leq T$.
The system is in the isothermal condition with inverse temperature $\beta$, and the possible states of the system are denoted as $w^1,w^2\cdots w^M$.
The transition from $w'$ to $w$ is written as $w' \to w$, where the transition rate at time $t$ is denoted by $P(w' \to w ;t)$.
We assume that for any states $w$ and any time $t$ there exists another state $w'$ such that $P(w'\to w ;t)\neq 0$.
The time evolution of the system is written as the master equation:
\begin{equation}
\frac{\partial P(w, t)}{\partial t}=J(w,t):= \sum_{w'} J(w' \to w;t),
\label{master}
\end{equation}
where $P (w, t)$ is the probability distribution of $w$ at time $t$, and $J(w'\to w;t) := P(w',t)P(w' \to w;t) -P(w,t)P(w \to w';t)$ represents the probability flux from $w'$ to $w$.

Jumps between states occur stochastically in each trajectory, where we denote the number of jumps by $N$.
The $i$-th jump occurs at time $t_i$ ($1\leq i\leq N$), and correspondingly the state changes from $w_{i-1}$ to $w_i$.
We write the initial and final time of the entire dynamics as $t_0:=0$ and $t_{N+1}:=T$.
The state at time $t$ is written as $w(t)$.
We define the total entropy production $\sigma _{\rm tot}$ as~\cite{FTreview}
\eq{
\sigma _{\rm tot}:=-\sum_{i=1}^N\beta Q(w_{i-1}\to w_i;t_i)+s(w(T),T)-s(w(0),0),
}
where
\eq{
Q (w' \to w; t)=-\frac{1}{\beta} \ln \(\frac{P(w' \to w;t)}{P(w \to w'; t)}\)
}
is the heat absorption by the system from the heat bath with the transition $w' \to w$ at time $t$, and $s(w,t):=-\ln P(w,t)$ represents the stochastic Shannon entropy at the state $w$ and time $t$.
Here, we assumed the local detailed balance condition~\cite{vanKampen}, and normalized the Boltzmann constant $k_{\rm B}$ to $1$.
The total entropy production characterizes the irreversibility, and satisfies the IFT $\la e^{-\sigma _{\rm tot}}\ra =1$ and the second law of thermodynamics $\la \sigma  _{\rm tot}\ra \geq 0$~\cite{Evans, Kurchan, Crooks, Jarzynski2}, where $\la \cdot \ra$ represents the ensemble average over all trajectories.

\subsection{General framework}\lb{s:gen-frame}

We now describe a general theoretical framework introduced in Ref.~\cite{SS}, which is applicable to both Szilard-type and autonomous demons in a unified way, as discussed in the subsequent subsections.
Under this framework, the entropy production for the total system is divided into contributions from subsets of individual transitions, where each {\it partial entropy production} satisfies an IFT.

We divide a set of all possible transitions $w^i\to w^j$ into two subsets: $\Omega$ and its complement $\Omega ^{\rm c}$.
In \ffref{f:restrict}(a), for example, the six red arrows represent the transitions included in $\Omega$ and the other sixteen black arrows represent the transitions included in $\Omega ^{\rm c}$. 
We divide the total entropy production $\sigma _{\rm tot}$ into the contributions from $\Omega$ and $\Omega ^{\rm c}$, which are denoted by $\sigma _\Omega$ and $\sigma _{\Omega ^{\rm c}}$, respectively.
Note that the subset $\Omega$ can be time-dependent in general.

\figin{10cm}{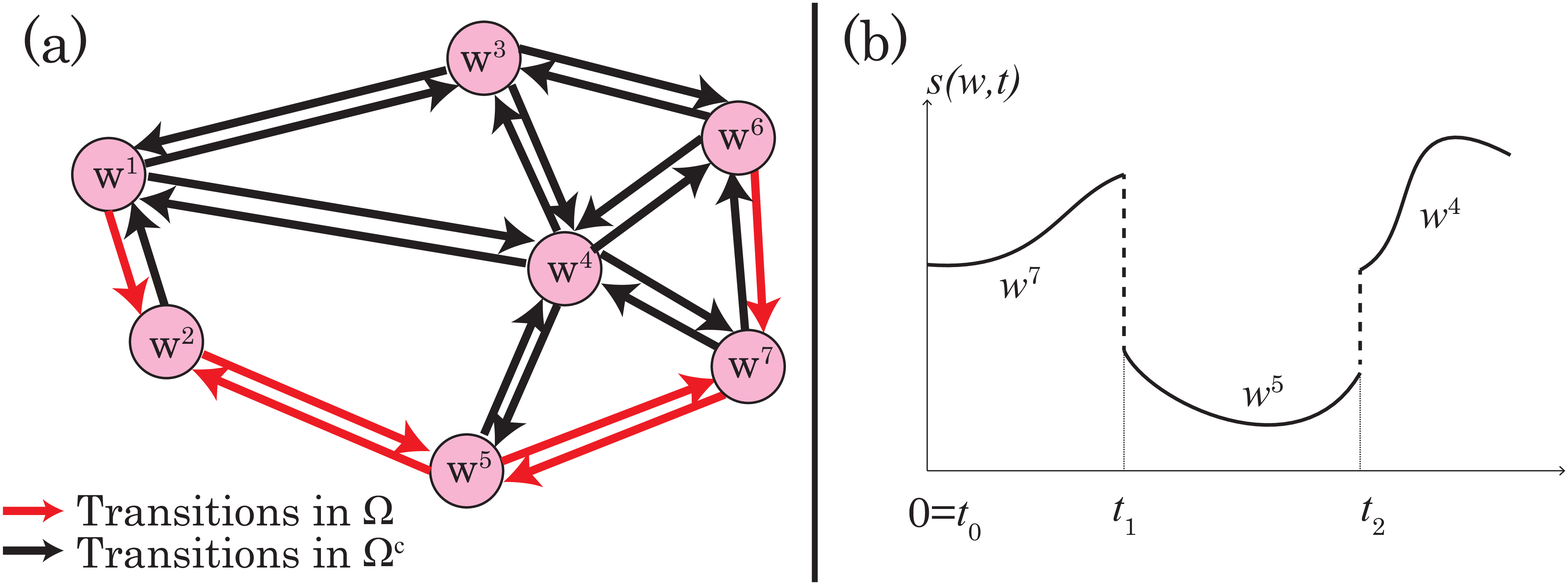}
{{\bf (a)}: An example of the choice of $\Omega$.
The six red arrows correspond to transitions in $\Omega$.
{\bf (b)}: A schematic of a time-series of the stochastic Shannon entropy along a single trajectory.
The initial state is $w^7$, and the jumps $w^7\to w^5$ and $w^5\to w^4$ occur at times $t_1$ and $t_2$, respectively.
The dashed lines represent the change in the stochastic Shannon entropy induced by the jumps, and the solid lines represent that induced by the change in the probability distribution.
}
{f:restrict}

First, we divide the probability flux at the state $w$ as $J(w,t)=J_{\Omega}(w,t)+J_{\Omega^{\rm c}}(w,t)$, where $J_{\Omega}(w,t)$ is defined as
\eqa{
J_{\Omega}(w,t):=\sum_{\{w'|(w'\to w)\in \Omega\}} J(w' \to w;t)
}{JOmega}
and $J_{\Omega^{\rm c}}(w,t)$ is defined in the same manner.
Using \eqqref{JOmega}, we define the partial entropy production $\sigma_\Omega$ as~\cite{SS}
\begin{equation}
\sigma_\Omega := -\beta Q_{\Omega} + \Delta s_{\Omega}.
\label{observable}
\end{equation} 
The first term of the right-hand side (rhs) of \eqqref{observable} represents the heat absorption accompanied by the transitions in $\Omega$:
\eq{
Q_{\Omega}:=\sum_{i=1}^{N} Q(w_{i-1} \to w_i;t_i) \delta_\Omega (w_{i-1}\to w_i ),
}
where $\delta_\Omega (w'\to w)$ takes $1$ if $(w'\to w) \in \Omega$ and takes $0$ otherwise.
The second term of the rhs of \eqqref{observable} represents the change in the stochastic entropy induced by the transitions in $\Omega$:
\eqa{
\Delta s_{\Omega}:=s_{\Omega ,{\rm jump}}-\int_0^T\frac{J_\Omega (w(t),t)}{P(w(t),t)}dt.
}{s_omega}
The first term of the rhs of \eqqref{s_omega} is defined as
\eqa{
s_{\Omega ,{\rm jump}}:=\sum_{i=1}^{N}\( s(w_{i},t_{i})-s(w_{i-1},t_{i})\) \delta_\Omega (w_{i-1}\to w_i),
}{somegajump}
which quantifies the contributions from jumps (dashed lines in \ffref{f:restrict}(b)) within $\Omega$.
The second term of the rhs of \eqqref{s_omega} evaluates the change induced by the probability flux (the solid lines in \ffref{f:restrict}(b)) within $\Omega$, which is confirmed by
\eqa{
\frac{\partial s(w,t)}{\partial t}=-\frac{1}{P(w,t)}\frac{\del}{\del t}P(w,t)=-\frac{ J_{\Omega}(w,t)}{P(w,t)} - \frac{J_{\Omega^{\rm c}}(w,t)}{ P(w,t)}.
}{J/Pdiv}

It is crucial that the partial entropy production $\sigma _\Omega$ satisfies
\eq{
\sigma _\Omega +\sigma _{\Omega ^{\rm c}}=\sigma  _{\rm tot},
}
and the IFT~\cite{SS}:
\eqa{
\la e^{-\sigma _\Omega}\ra =1.
}{restrict}
Applying Jensen's inequality to \eqqref{restrict}, we obtain 
\eqa{
\la \sigma _\Omega\ra \geq 0.
}{partial-ineq}
In the following two subsections, we show two IFTs for Szilard-type demons \eqref{SU} and that for autonomous demons \eqref{class2eq} which are particular cases of \eqqref{restrict}.

\subsection{IFT for the Szilard-type demon}\lb{s:SU}

\figin{7cm}{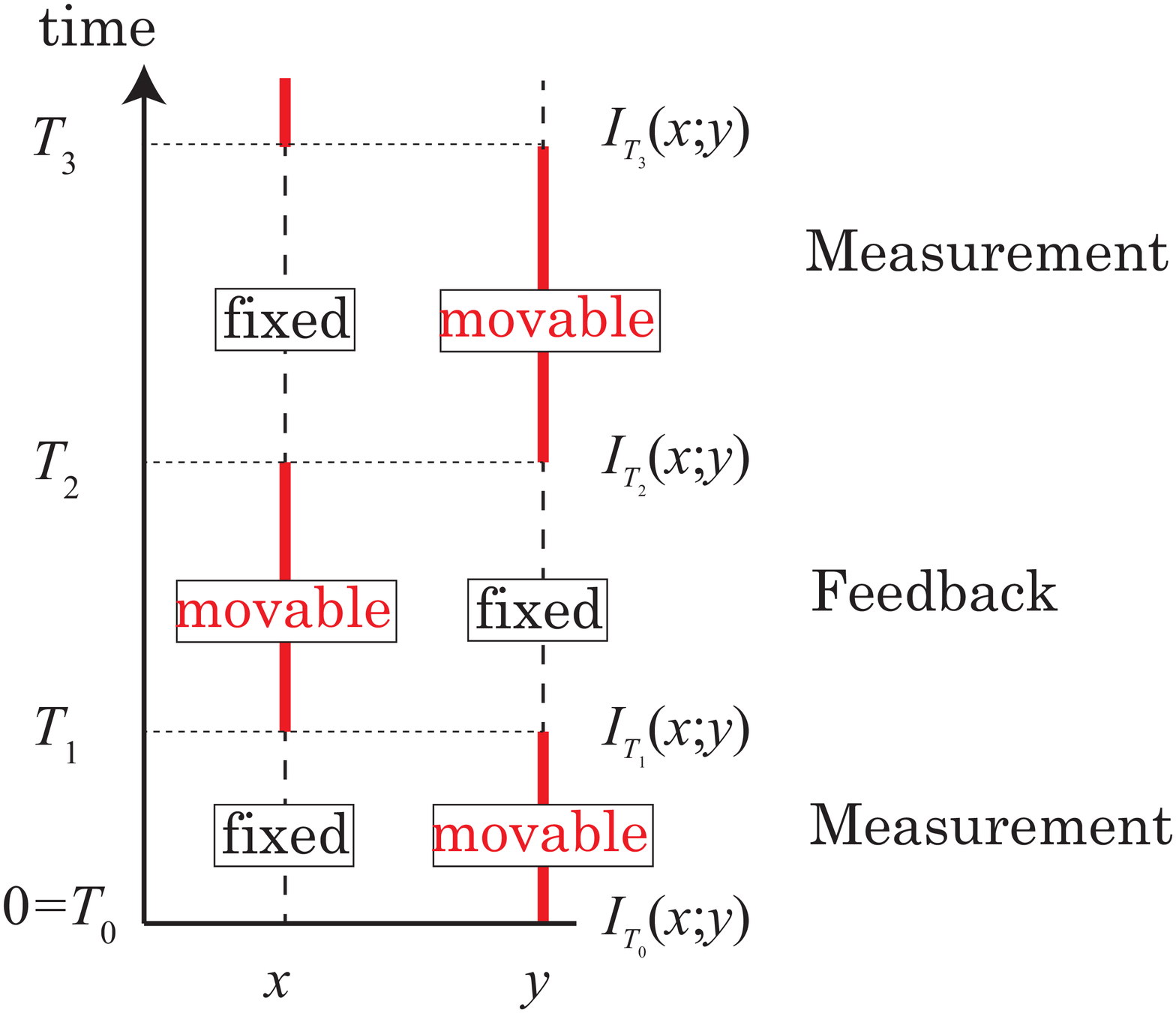}
{Repeated measurement and feedback processes.
The dashed lines indicate frozen phases and the red bold lines indicate movable phases.
}{f:rep-SU}

In this and the next subsections, we consider a bipartite system that consists of subsystems $X$ and $Y$, where the state is written as $w=(x,y)$, with the transition rates satisfying $P((x,y)\to (x',y');t)=0$ for $x\neq x'$ and $y\neq y'$.
We define the entropy production of subsystem $X$ as 
\eqa{
\sigma _X:=-\beta Q_X+\Delta s_X,
}{sigmax}
where $Q_X$ is the heat absorbed by subsystem $X$, and $\Delta s_X:=-\ln P(x(T),T)+\ln P(x(0),0)$ is the change in the stochastic Shannon entropy of only $x$.
In contrast to the averaged entropy production of the total system $\la \sigma _{\rm tot}\ra$, that of the subsystem $\la \sigma _X \ra$ can be negative.
In the case of the Szilard engine~\cite{Szilard}, which is a composite system of an engine and a memory, the entropy production of the engine is negative on average corresponding to the extraction of work from the isothermal cycle.

In previous works~\cite{SU2012,SU2012full}, it has been discussed that modified IFT and second law still hold under these setups, if the change in the correlation between the engine and the memory is taken into account.
In the case of the Szilard engine, the measurement process increases the correlation, and the feedback process decreases the correlation.
In general, the correlation is quantified by the mutual information.
The stochastic mutual information between the state of the engine $X$ and the state of the memory $Y$ is defined as
\eq{
I_t(x;y):=\ln \frac{P(x,y,t)}{P(x,t)P(y,t)},
}
whose ensemble average 
\eq{
\la I_t(x;y)\ra =\sum_{x,y}P(x,y,t)\ln \frac{P(x,y,t)}{P(x,t)P(y,t)}
}
is the mutual information~\cite{Cover-Thomas}.
The mutual information is zero if there is no correlation, and becomes larger if the correlation is stronger.

\figin{7cm}{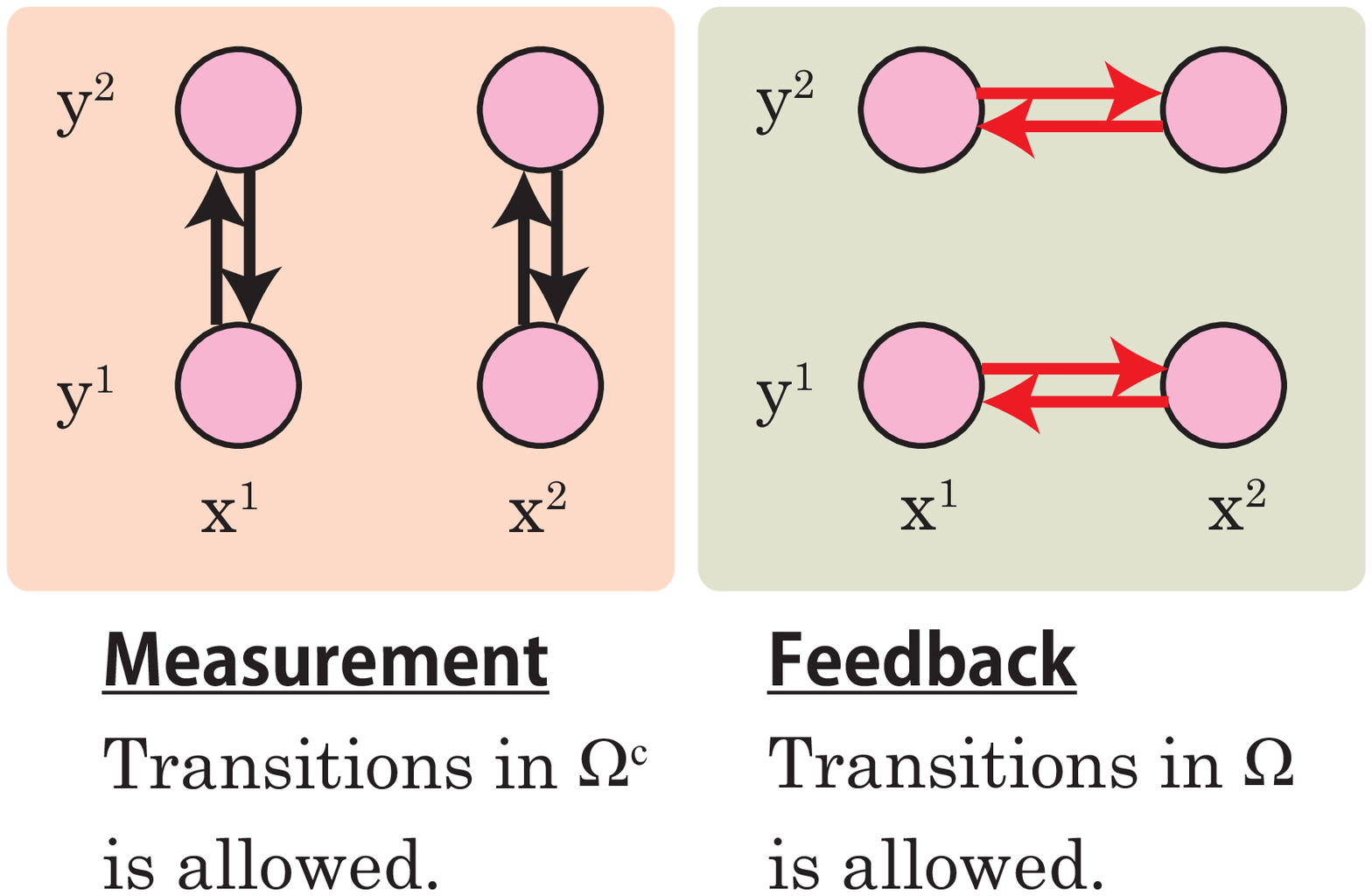}
{The choice of $\Omega$ in the proof of the IFT for Szilard-type demons \eqref{SU}.
}
{f:SU-arrow}

In order to generalize the setup of the Szilard engine, we here suppose that there are $m$-times of measurement and feedback processes in a composite system of $X$ and $Y$ during the time interval from $T_0=0$ to $T_{2m}$.
During the time interval $T_{2n}\leq t< T_{2n+1}$ ($T_{2n+1}\leq t<T_{2n+2}$) with $0\leq n\leq m-1$, only $Y$ ($X$) can evolve and $X$ ($Y$) is fixed (see \ffref{f:rep-SU}), corresponding to the measurement (feedback) phase.
These conditions are written as
\balign{
P(x\to x';y,t)&=0 \ \ (T_{2n}\leq t< T_{2n+1}), \lb{SUset1} \\
P(y\to y';x,t)&=0 \ \ (T_{2n+1}\leq t< T_{2n+2}), \lb{SUset2}
}
where we abbreviated the transition rates as $P(x\to x';y,t):=P((x,y)\to (x',y);t)$ and $P(y\to y';x,t):=P((x,y)\to (x,y');t)$.
This setup represents the repeated measurement and feedback processes between the engine $X$ and the memory $Y$.
Here, by applying \eqqref{restrict} with setting all transitions in $X$ as $\Omega$ (see \ffref{f:SU-arrow}), the entropy production of the engine $\sigma _X$ satisfies the following IFT and the generalized second law~\cite{SU2012full}:
\balign{
\la e^{-\sigma _X+\sum_n\Delta I_{X,n}}\ra &=1, \lb{SU} \\
\la \sigma _X\ra -\sum_n\la \Delta I_{X,n}\ra &\geq 0,
}
where $\Delta I_{X,n}:=I_{T_{2n+2}}(x;y)-I_{T_{2n+1}}(x;y)$ represents the change in the mutual information between before and after the $n$-th feedback process.
This IFT implies that if we consume the correlation between the engine and the memory (i.e., $\sum_n\la \Delta I_{X,n}\ra<0$), the entropy production in the engine can be negative (i.e., $\la \sigma _X\ra <0$) up to the amount of the consumption.
In contrast, if we establish their correlation, we need the corresponding extra cost.
Equality \eqref{SU} is a special case of a more general IFT that is discussed below.

\subsection{IFT for general measurement and feedback processes }\lb{class2-rel}

In this subsection, we mention on the case of the general measurement and feedback processes with bipartite systems, which is applicable for example to autonomous information processing in biological systems~\cite{Tostevin, Lan}.
The transition rate of $X$ depends on the state of memory $Y$, and vice versa, although the time-separation of the measurement and feedback [Eqs.~\eqref{SUset1} and \eqref{SUset2}] is not assumed.

\figin{6cm}{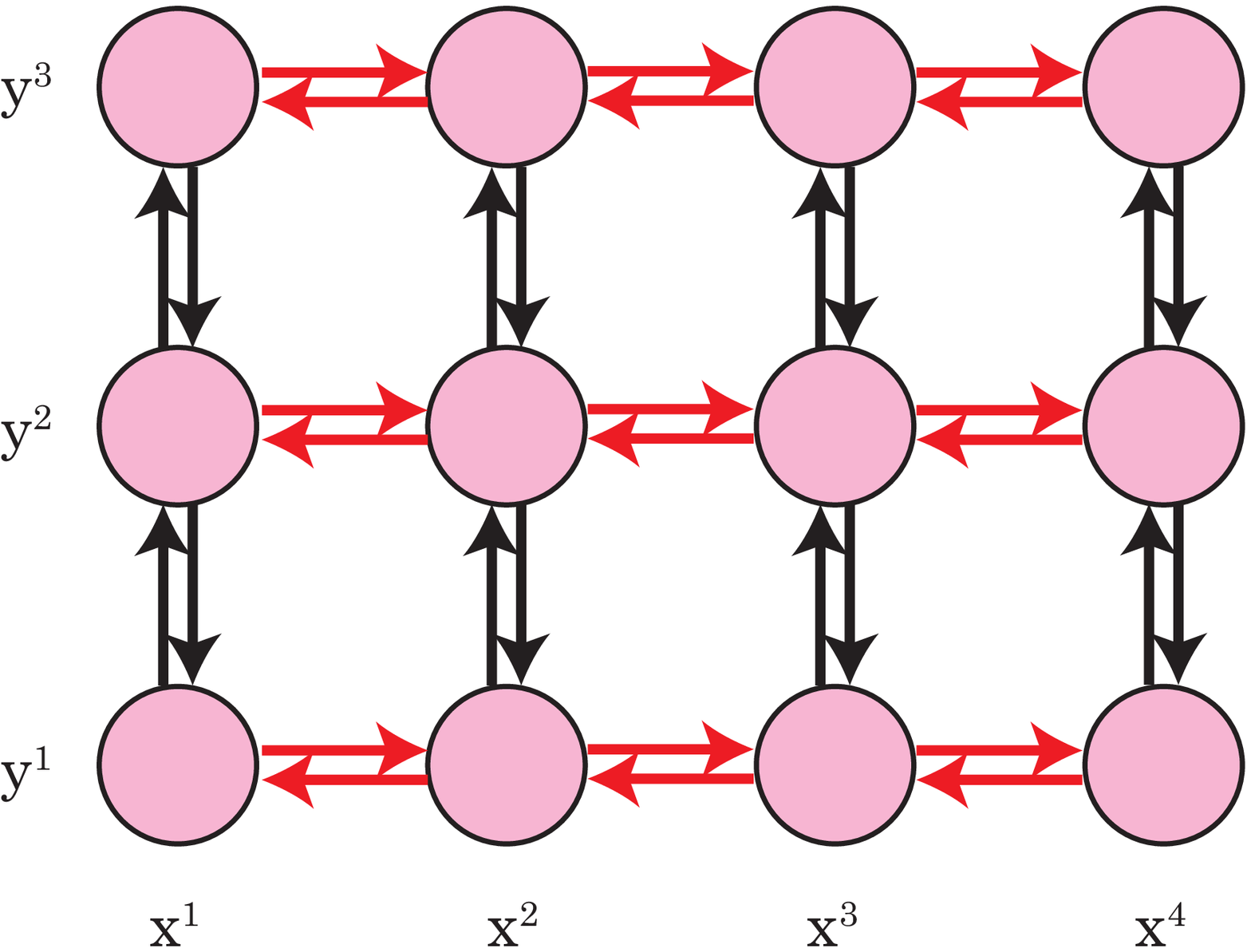}
{The choice of $\Omega$ in the proof of the IFT for general measurement and feedback processes \eqref{class2eq}.
}{f:class2}

We define the entropy production associated only with $X$ by the same form as \eqqref{sigmax}.
By applying \eqqref{restrict} with setting all transitions in $X$ as $\Omega$ (see \ffref{f:class2}), it has been shown that the following IFT holds~\cite{SS}:
\eqa{
\la e^{-\sigma _X+\Delta I_X}\ra =1,
}{class2eq}
where $\Delta I_X$ represents the change in the mutual information contributed from $X$~\cite{Liang, Majda, Allahverdyan, Jordan2014} defined as
\eq{
\Delta I_X:= I_{X,{\rm jump}}+\int_0^TF_X(x(t),y(t),t)dt.
}
The first term $I_{X,{\rm jump}}$ represents the change in the mutual information with realized jumps of $X$ defined as
\eq{
I_{X,{\rm jump}}:=\sum_{i=1}^N\( I_{t_i}(x_i;y_i)-I_{t_{i}}(x_{i-1};y_{i-1})\) \delta_{y_i,y_{i-1}} ,
}
where $\delta$ is the Kronecker delta.
Here, we used notation $w_i=:(x_i, y_i)$.
The second term involves a time integral over the entire time interval; its integrand $F_X(x,y,t)$ represents the change in the mutual information induced by the probability flux of $X$, which is defined as
\eqa{
F_X(x,y,t):=\frac{1}{P(x,y,t)}J_X(x,y,t)-\frac{1}{P(x,t)}J_X(x,t),
}{Fxdef}
where $J_X(x,t):=\sum_yJ_X(x,y,t)$.

\section{Autonomous Maxwell's demon with separated measurement and feedback process}\lb{s:class1}

In contrast to the IFT for the Szilard-type demons \eqref{SU}, the IFT for general measurement and feedback \eqref{class2eq} includes a time-integral term over the entire time interval.
The most important difference between these setups is whether the time interval of the measurement phase and the feedback phase are separated externally.
Although autonomous systems are not controlled externally,  two phases can be separated even in autonomous systems by introducing an additional stochastic variable which determines whether the present state is in the measurement phase or in the feedback phase.
To clarify this point, we construct an autonomous model with separated measurement and feedback phases, and derive an IFT satisfied in this system.

\subsection{Model and setups}\label{s:8state}

\figin{15cm}{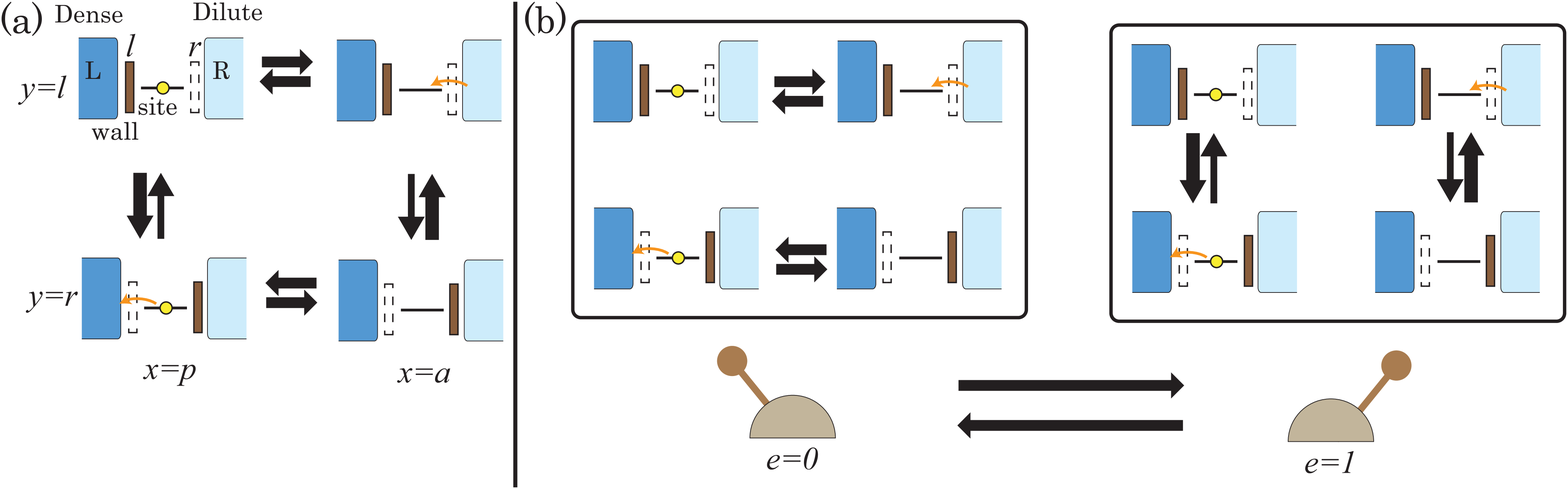}
{{\bf (a)}: The state space of the 4-state model.
If a particle is (is not) in the site, the wall tends to move right (left).
{\bf (b)}: The state space of the 8-state model.
For $e=1(0)$, only the state of the wall (the particle) can change.
The additional variable $e$ also changes stochastically.
}{f:8state}

Let us start from simple models with and without measurement-feedback separation.
We first consider the autonomous 4-state model introduced in Ref.~\cite{SS}, where the measurement phase and the feedback phase are not separated. 
The 4-state model is a bipartite system which transports particles from a dense particle bath L to a dilute particle bath R (see Fig.~\ref{f:8state}(a)).
There is a site between two baths with at most one particle, where $x\in\{p,a\}$ represents the presence $p$ or the absence $a$ of a particle at the site.
We then introduce a wall which prohibits jumps of particles.
If the wall is between the site and bath L (R), we denote the state of the wall as $y=l \ (r)$.
The position of the wall tends to be $r$ ($l$) if a particle exists (does not exist) at the site.
With these processes, particles are transported against the chemical potential gradient.

Now we construct an autonomous model named the 8-state model with the measurement-feedback separation, by adding a new stochastic variable $e\in \{0,1\}$ to the 4-state model (see \ffref{f:8state}(b)).
The role of $e$ is to determine whether the present state is in the measurement phase or the feedback phase.
When $e=0 \ (1)$, only particles (the wall) can move and the wall (particles) is fixed, which corresponds to the feedback (measurement) phase.
Since the probability distributions and transition rates are independent of time, we omit the argument of $t$. 
These conditions are written as $P(l\to r;x,0)=P(r\to l;x,0)=0$, $P(p\to a;y,1)=P(a\to p;y,1)=0$.
We assume that the transition rates of $e$ are independent of $x$ and $y$, and write $P_{01}:=P(0\to 1;x,y)$ and $P_{10}=P(1\to 0;x,y)$.

We can easily generalize this setup to the case with more than eight states.
Suppose a composite system with three variables; $x$, $y$, and an additional variable $e\in \{ 0,1\}$, which is in the stationary distribution with time-independent transition rates.
We assume that only one of $\{ x,y,e\}$ changes within a single transition. 
The transition rates also satisfy the following conditions (see also \ffref{f:class1}(a)):
\balign{
P(e\to e';x,y)&=P_{ee'}, \lb{NCCIC} \\
P(x\to x';y,1)&=0, \\ 
P(y\to y';x,0)&=0.
}

\subsection{IFT for separated autonomous demons}\lb{s:class1-set}

We now discuss an IFT for autonomous demons with separated measurement and feedback (e.g., the 8-state model).
The entropy production of engine $X$ is defined as 
\balign{
\sigma _X&:=-\beta Q_X+\Delta s_X \nt \\
&:=\sum_{i=1}^NQ(x_{i-1},y_{i-1},e_{i-1}\to x_{i},y_{i},e_{i})\delta _{y_{i-1},y_{i}}\delta _{e_{i-1},e_{i}}+s(x_N)-s(x_0).
}
By using the stationary distribution $P_{\rm ss}(x):=\int dyde P_{\rm ss}(x,y,e)$, we defined the stochastic entropy of $x$ as $s(x):=-\ln P_{\rm ss}(x)$.
Under the setup described in Sec.~\ref{s:8state}, $\sigma _X$ satisfies the following IFT:
\eqa{
\la e^{-\sigma _X+\sum I_{0,i}^{\rm ch}+I^0(x_N;y_N)-I^0(x_0;y_0)}\ra=1.
}{class1eq}
Here, $I_{0,i}^{\rm ch}$ is defined as
\eqa{
I^{\rm ch}_{0,i}:=I(x_{i-1};y_{i-1}|e_{i-1})(\delta _{e_{i-1},0}-\delta _{e_{i},0}),
}{ench-def}
which subtracts (adds) the mutual information at $e=1 \ (0)$ when the state changes from the measurement (feedback) phase $e=1 \ (0)$ to the feedback (measurement) phase $e=0 \ (1)$.
The conditional mutual information is defined as
\eq{
I(x;y|e):=\ln \frac{P_{\rm ss}(x,y|e)}{P_{\rm ss}(x|e)P_{\rm ss}(y|e)},
}
where $P_{\rm ss}(x|e):=\int dy P_{\rm ss}(x,y,e)/P_{\rm ss}(e)$.
The definition of  $\Delta I_{X,n}$ in \eqqref{SU} is regarded as the change in the mutual information between the endpoint of the measurement processes and the endpoint of the feedback processes. 
$I^0(x;y)$ is defined as $I^0(x;y):=I(x;y|e)\delta _{e,0}$, which counts the mutual information at the initial and the final states if these states are in $e=0$.

\figin{14cm}{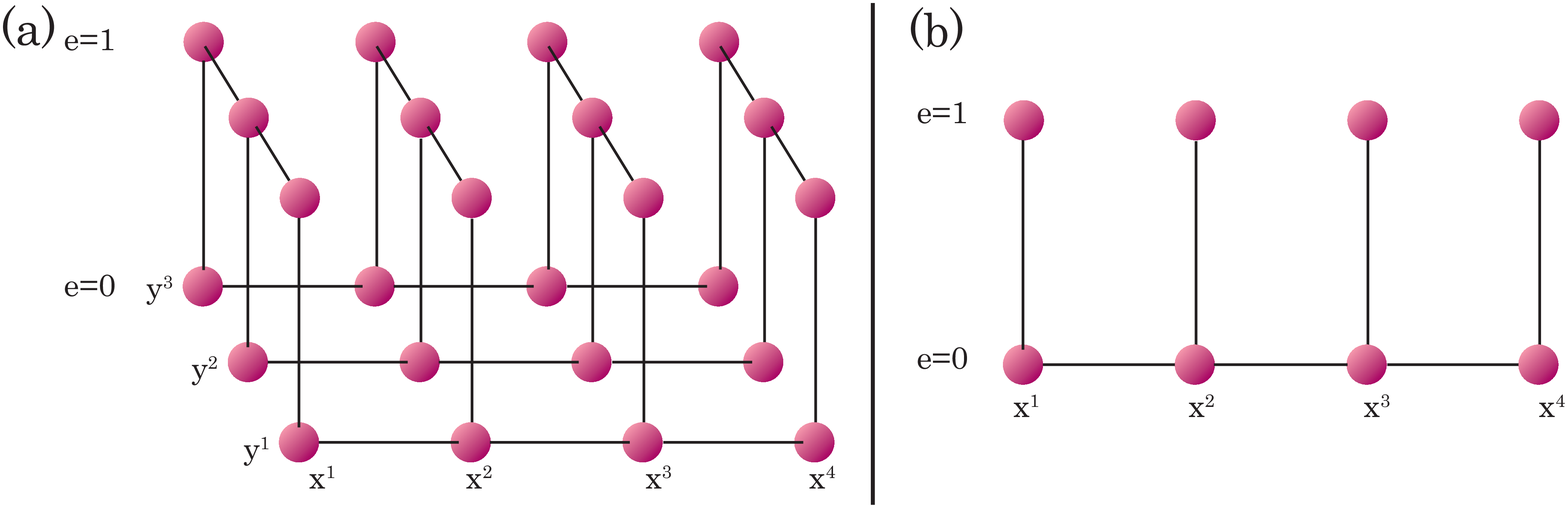}
{{\bf (a)}: The state space of a system which satisfies the setup in Sec.~\ref{s:8state}.
For $e=1$, only $y$ can change, and for $e=0$, only $x$ can change.
{\bf (b)}: The same state space seen along $y$ axis.
Since $x$ does not change for $e=1$, $P_{\rm ss}(x|1)=P_{\rm ss}(x|0)=P_{\rm ss}(x)$ holds in stationary state.}
{f:class1}

We now derive \eqqref{class1eq} from \eqqref{restrict} by setting $\Omega$ to $\{ (x',y',e')\to (x,y,e)|e=0\}$ (see also \ffref{f:8-arrow}).
In this case, since the system is in the stationary state, $\Delta s_\Omega$ is calculated as
\balign{
\Delta s_\Omega =&\sum_{i=1}^{N}s(x_{i-1},y_{i-1},e_{i-1})(\delta _{e_{i-1},0}-\delta _{e_{i},0}) \nt \\
&+s(x_{N},y_{N},e_{N})\delta _{e_{N},0}-s(x_0,y_0,e_0)\delta _{e_0,0}, 
}
where we used $J_\Omega (x,y,e)=0$ for all $(x,y,e)$.
The difference between $Q_{\Omega}$ and $Q_X$ is equal to the heat absorption accompanying the transitions from $e=1$ to $e=0$:
\balign{
\beta Q_{\Omega}-\beta Q_X&=\sum_{i=1}^N\ln \frac{P_{\rm ss}(1)}{P_{\rm ss}(0)}\delta _{e_i,0}\delta _{e_{i-1},1} \nt \\
&=\sum_{i=1}^N\ln P_{\rm ss}(e_{i-1})(\delta _{e_{i},0}-\delta _{e_{i-1},0})+\ln P_{\rm ss}(0)(\delta _{e_0,0}-\delta _{e_N,0}), 
}
where we used the stationary condition for $e$ such that $P_{\rm ss}(0)P_{01}=P_{\rm ss}(1)P_{10}$.
Since $y$ is fixed while $e=0$ and $x$ is fixed while $e=1$, it is easy to show that
\balign{
&\sum_{i=1}^N\( s(x_{i-1})+s(y_{i-1})\) (\delta _{e_{i-1},0}-\delta _{e_{i},0}) \nt \\
=&s(x_N)\delta _{e_N,1}-s(y_N)\delta _{e_N,0}-s(x_0)\delta _{e_0,1}+s(y_0)\delta _{e_0,0}.
}
We also note that
\balign{
P_{\rm ss}(x|1)&=P_{\rm ss}(x|0)&=P_{\rm ss}(x) \lb{yaxis} \\
P_{\rm ss}(y|1)&=P_{\rm ss}(y|0)&=P_{\rm ss}(y) 
}
for all $x, y$ (see \ffref{f:class1}(b)), which follows from the fact that $P_{01}$ and $P_{10}$ are constant and independent of $(x,y)$.
Finally, we arrive at
\balign{
\fl
\sigma _{\Omega}
&=&-\beta Q_\Omega +\sum_{i=1}^{N}\( s(x_{i-1},y_{i-1},e_{i-1})-s(x_{i-1})-s(y_{i-1})-s(e_{i-1})\) (\delta _{e_{i-1},0}-\delta _{e_{i},0}) \nt \\
\fl&&+(s(x_{N},y_{N},e_{N})-s(y_N)+\ln P(0))\delta _{e_{N},0}+s(x_N)\delta _{e_N,1} \nt \\
\fl&&-(s(x_0,y_0,e_0)-s(y_0)+\ln P(0))\delta _{e_0,0}-s(x_0)\delta _{e_0,1} \nt \\
\fl&=&-\beta Q_X+s(x_N)-s(x_0)-\sum_{i=1}^N I_{0,i}^{\rm ch}-I^0(x_N;y_N)+I^0(x_0;y_0),
}
which implies \eqqref{class1eq}.

\figin{6cm}{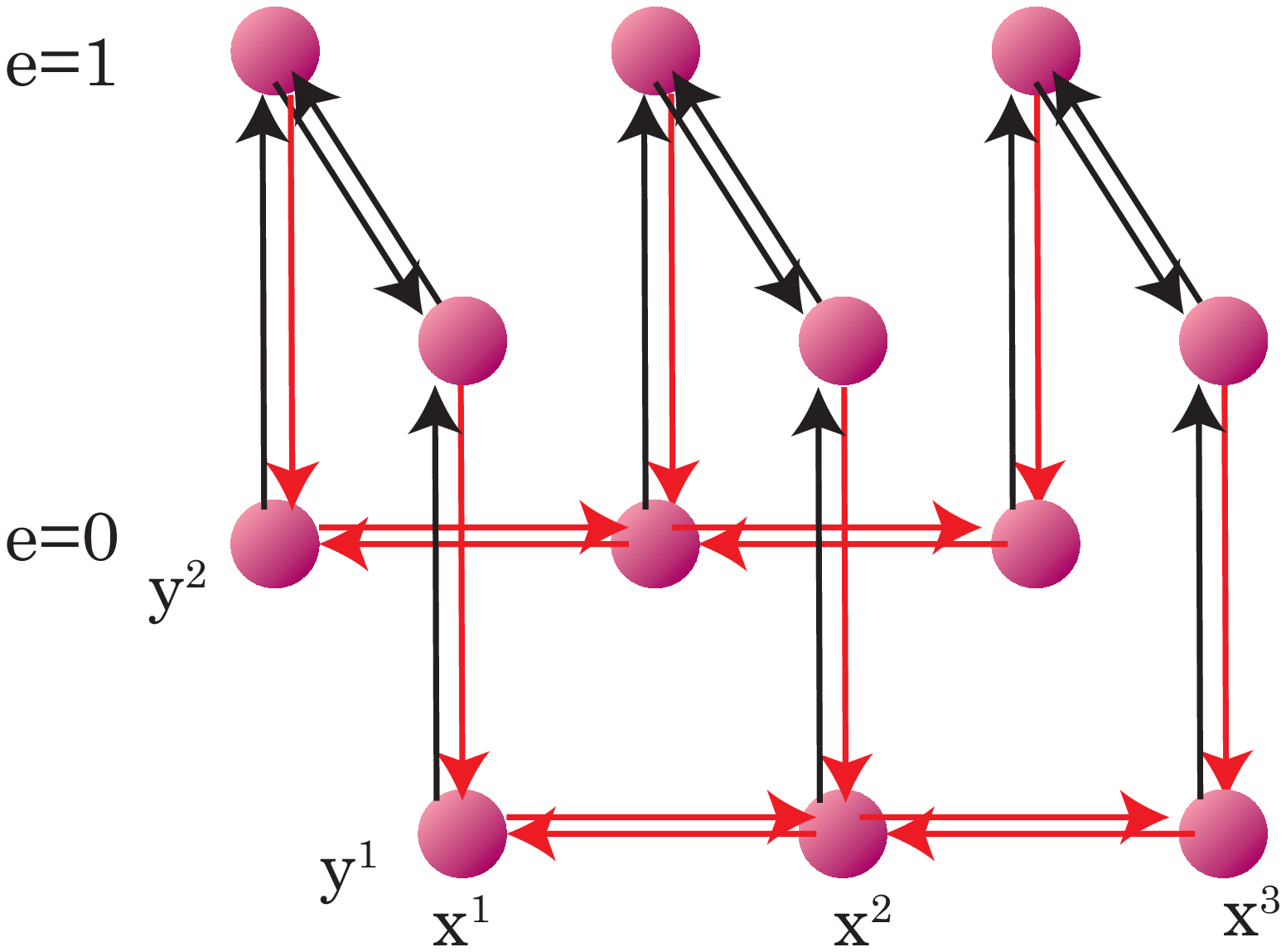}
{The choice of $\Omega$ in the proof of \eqqref{class1eq}.
Transitions $\{ (x',y',e')\to (x,y,e)|e=0\}$ are set as $\Omega$, which are colored by red.
}{f:8-arrow}

\subsection{State-space reduction of autonomous demon}\lb{s:redu}

In this subsection, we show that the IFT for non-separated autonomous demons \eqref{class2eq} in stationary states can be naturally derived from the IFT for separated autonomous demons \eqref{class1eq}, by considering the reduction of the variable $e$. 
Suppose that there are two models, $A$ and $B$.
Model $A$ has three variables $(x,y,e)$, and satisfies the setup described in Sec.~\ref{s:8state}.
Model $B$ has two variables $(x,y)$, and satisfies the setup described in Sec.~\ref{class2-rel}.
Hence, \eqqref{class1eq} holds in model $A$ and \eqqref{class2eq} holds in model $B$.
In the following, superscript $A$ ($B$) represents quantities in model $A$ ($B$).
By using the transition rates of model $A$, we set the transition rates of model $B$ to
\balign{
P^B(x\to x';y)&=P_{\rm ss}^A(0)P^A(x\to x';y,0), \\
P^B(y\to y';x)&=P_{\rm ss}^A(1)P^A(y\to y';x,1).
} 
Here, we introduce a quantity $k$:
\eq{
k:=P_{01}P_{\rm ss}^A(0)=P_{10}P_{\rm ss}^A(1),
}
which characterizes the typical rate of the transition in $e$.
By taking the limit $k \to \infty$ in model $A$ with fixed $P^A(x\to x';y,e)$, $P^A(y\to y';x,e)$, $P_{\rm ss}^A(0)$, and $P_{\rm ss}^A(1)$, we obtain a reduced Markovian dynamics with the variables $(x,y)$, which is equivalent to the dynamics of model $B$.

\figin{12cm}{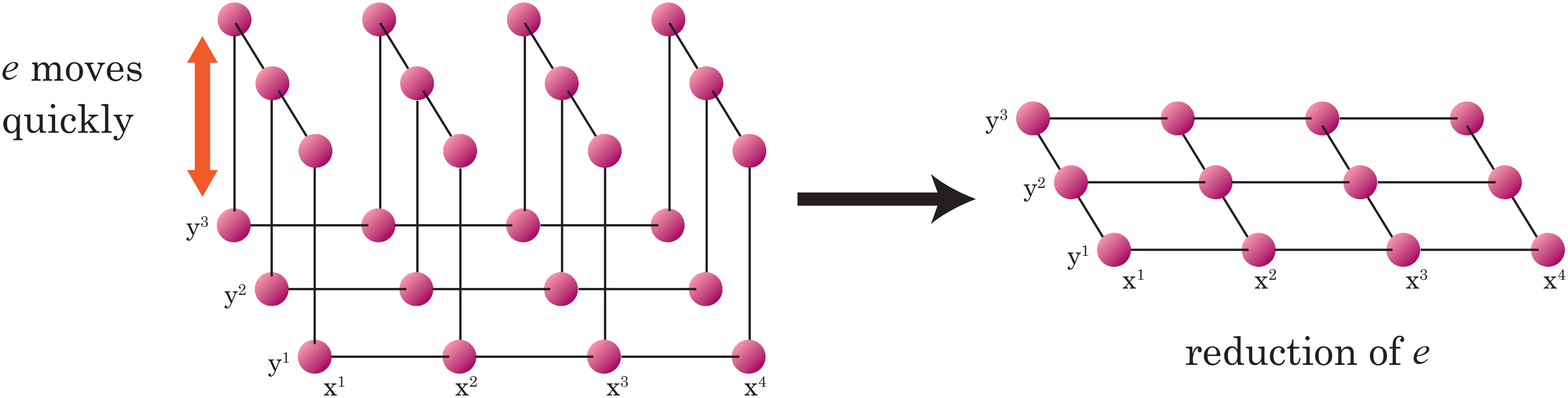}
{In the setup of Sec.~\ref{s:8state}, we take the limit that $e$ changes quickly, and obtain the reduced model with two variables $x$ and $y$.}
{f:contra}

Now we compare \eqqref{class1eq} for model $A$ in the limit $k \to \infty$ and \eqqref{class2eq} for model $B$.
For sufficiently large $k$, the number of back-and-forth transitions between $e=0$ and $e=1$ is $k +O(\sqrt{k})$ per unit time, where the second term represents the fluctuation.
Then, the total count of $\sum I_{0,i}^{\rm ch}$ per unit time during the stay at $(x^a,y^b)$ is calculated as
\balign{
\fl&\( k +O(\sqrt{k})\)  \( I^A(x^a,y^b|0)-I^A(x^a,y^b|1)\) \nt \\
\fl=&\( k +O(\sqrt{k})\) \ln \frac{P_{\rm ss}^A(x^a,y^b|0)}{P_{\rm ss}^A(x^a,y^b|1)} \nt \\
\fl=&\( k +O(\sqrt{k})\) \( \frac{P_{\rm ss}^A(x^a,y^b|0)}{P_{\rm ss}^A(x^a,y^b|1)}-1+O\(\frac{1}{k ^2}\)\) \nt \\
\fl=&\frac{\sum_{x^c}P_{\rm ss}^A(x^c,y^b,0)P^A(x^c\to x^a;y^b,0)-P_{\rm ss}^A(x^a,y^b,0)P^A(x^a\to x^c;y^b,0)}{P_{\rm ss}^A(x^a,y^b|1)}+O\(\frac{1}{\sqrt{k}}\) \nt \\
\fl=&\frac{\sum_{x^c}P_{\rm ss}^B(x^c,y^b)P^B(x^c\to x^a;y^b)-P_{\rm ss}^B(x^a,y^b)P^B(x^a\to x^c;y^b)}{P_{\rm ss}^B(x^a,y^b)}+O\(\frac{1}{\sqrt{k}}\) \nt \\
\fl=&\frac{J_X^B(x^a,y^b)}{P_{\rm ss}^B(x^a,y^b)}+O\(\frac{1}{\sqrt{k}}\) \nt \\
\fl=&F_X(x^a,y^b,t)+O\(\frac{1}{\sqrt{k}}\) . \label{class1contra}
}
In the second line, we used \eqqref{yaxis}.
In the third and fifth lines, we used
\eq{
P_{\rm ss}^A(x^a,y^b|e)=P_{\rm ss}^A(x^a,y^b)+O\( \frac{1}{k}\) =P_{\rm ss}^B(x^a,y^b)+O\( \frac{1}{k}\) ,
}
for $e=0,1$.
In the fourth line, we used the stationary condition at $(x^a,y^b,0)$:
\balign{
\fl&\frac{k}{P_{\rm ss}^A(0)}\cdot P_{\rm ss}^A(x^a,y^b,0)-\frac{k}{P_{\rm ss}^A(1)}\cdot P_{\rm ss}^A(x^a,y^b,1) \nt \\
\fl=&\sum_{x^c}\( P_{\rm ss}^A(x^c,y^b,0)P^A(x^c\to x^a;y^b,0)-P_{\rm ss}^A(x^a,y^b,0)P^A(x^a\to x^c;y^b,0)\) . 
}
In the last line, we used the fact that the second term of the rhs of \eqqref{Fxdef} is zero in the stationary state due to
\eq{
J_X(x,t)=\frac{\del}{\del t}P(x,t)=0.
}
From \eqref{class1contra}, it is straightforward to find that \eqqref{class2eq} for model $B$ is equivalent to \eqqref{class1eq} for model $A$ in the limit $k\to \infty$.
This result indicates that the time-integral term in \eqqref{class2eq} appears due to the measurement phase and the feedback phase being unseparated in the setup of Sec.~\ref{class2-rel}

\section{Role of separated measurement and feedback in IFTs}\lb{s:class1-class2}

We discussed in the previous section that the IFT \eqref{class1eq} satisfied in our separated autonomous demon model is similar to that of the Szilard-type systems \eqref{SU}. 
In contrast, the IFT \eqref{class2eq} for the general measurement and feedback systems contained a time-integral term $F_X(x,y,t)$, which we found an interpretation as the mutual information flow in the fast switching limit as shown in \eqqref{class1contra}. 
We here aim to clarify the relation between the existence of the time-integral term in the IFT and the separation of measurement and feedback dynamics.

Let us first recall how we set $\Omega$ when we derived the IFTs, from the viewpoint of the general framework in Sec.~\ref{s:gen-frame}.
First, we define a set of transitions whose final state is $w$ as $S_w:=\{ (w'\to w)| w'\in W, (w'\to w)\in G \}$, where $W$ represents the set of all possible states and $G$ represents the set of all possible transitions.
Then, the common feature of the choices of $\Omega$ for \eqqref{SU} and \eqqref{class1eq} is that $S_w$ satisfies $S_w\subseteq \Omega$ or $S_w\subseteq \Omega ^{\rm c}$ for any $w$.
In the case of \eqqref{SU}, $S_w$ is time-dependent, and all $S_w$ satisfy $S_w \subseteq \Omega$ ($\Omega ^{\rm c}$) in the feedback (measurement) phase (see Fig.~\ref{f:SU-arrow}).
In the case of \eqqref{class1eq}, for $w$ with $e=0$ ($e=1$), $S_w$ satisfies $S_w \subseteq \Omega$ ($\Omega ^{\rm c}$) (see Fig.~\ref{f:8-arrow}).
This property implies the separation of the measurement phase and the feedback phase.
We call this condition the {\it separation condition}.
Conversely, the measurement phase and the feedback phase are unseparated, when we can find some $(w,w',w'')$ such that $(w' \to w) \in \Omega$ and $(w'' \to w) \in \Omega^{\rm c}$ (see \ffref{f:class2}), as in the case of the general measurement and feedback setup including the reduced dynamics discussed in Sec.~\ref{s:redu}.

\figin{5cm}{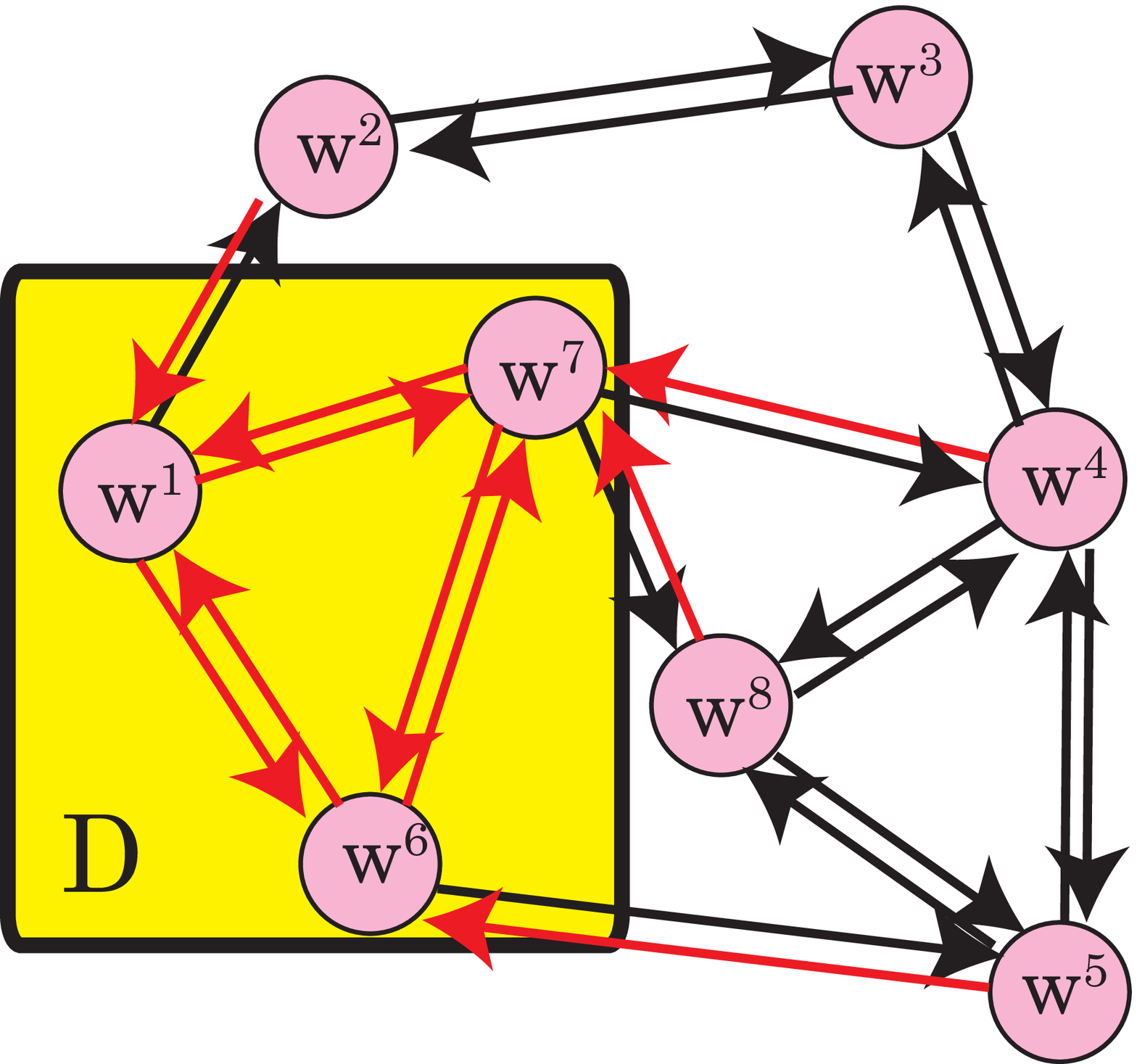}
{An example of the division of the state space.
The subset of states $D$ is $\{ w^1,w^6,w^7\}$.
We set the ten red arrows to $\Omega$ in this case.}
{f:restrict-state}

To confirm that the separation condition plays a crucial role for the time-integral term in IFT, we show another IFT for the division in a state space.
Corresponding to a subset $\Omega$ with separation condition, we divide the possible states into two groups (see \ffref{f:restrict-state}): 
\begin{eqnarray}
D &:=& \{ w |S_w\subseteq \Omega \} \label{sep1} \\
D^{\rm c}  &:=& \{ w | S_w \subseteq \Omega ^{\rm c}  \}. \label{sep2}
\end{eqnarray}
Owing to the separation condition, $D^{\rm c}$ is indeed the complement of $D$.
Although $D$ is time-dependent in general, in the following we assume that $D$ is time-independent for simplicity.
We now show that $\sigma _{\Omega}$ can be transformed into a form without any time-integral term.
By focusing on the change in the stochastic Shannon entropy, \eqqref{s_omega}, the integrand of the time-integral term can be written as
\eq{
\frac{J_{\Omega _D}(w,t)}{P(w,t)}=
\lb{f-in-state}
\cases{
\frac{\del}{\del t}\ln P(w,t) & $w\in D$ \\
0 & otherwise, \\
}}
which is the crucial property of the separation condition.
Adding the contribution from the jumps, we have
\begin{eqnarray}
\fl \Delta s_{\Omega} &=& s_{\Omega,{\rm jump}} - \int _0 ^T \frac{J_{\Omega} (w,t)}{P(w,t)} \nt \\
\fl &=&  \sum_{i=1} ^{N} (s(w_i,t_i)-s(w_{i-1},t_i)) \delta_{D}(w_i) + \sum_{i=0} ^{N} (s(w_{i},t_{i+1}) -s(w_{i},t_{i}) ) \delta_{D}(w_i) \nt \\
\fl &=&  \sum_{i=1}^N s^{\rm ch}_{D,i} + s_D(w_N,T)-  s_D(w_0,0). \label{sepStates}
\end{eqnarray}
where $\delta _D(w)$ takes 1 if $w\in D$, and 0 otherwise.
Note that $\delta _{\Omega}(w'\to w)=\delta _D(w)$ holds for $\Omega$ with the separation condition.
We also introduced the entropy exchange between $D$ and $D^{\rm c}$ through the $i$-th transition
\eq{
s^{\rm ch}_{D,i}:=
\cases{
s(w_{i-1},t_i) & $w_{i-1}\in D$, $w_i\notin D$ \\
-s(w_{i-1},t_i) & $w_{i-1}\notin D$, $w_i\in D$ \\
0 & otherwise,}
}
and the entropy associated with $D$:
\eq{
s_D(w,t):=s(w,t)\delta _D(w).
}
The entropy exchange $s^{\rm ch}_{D,i}$ represents the initial and the final entropy of the dynamics in $D$, which corresponds to $I_{0,i}^{\rm ch}$ in \eqqref{class1eq}.
Therefore, the IFT for the case of divided states does not include the time-integral term.
We note that the contribution from the entropy exchange in Eq.~(\ref{sepStates}) corresponds to the change in the mutual information, which appeared in the IFTs for Szilard-type systems and the model with separated measurement and feedback phases.

\section{Conclusion}\lb{s:conclusion}

In this paper, we clarified the difference between the Szilard-type demons and autonomous bipartite demons.
By introducing another type of autonomous demon, in which the measurement phase and the feedback phase are separated, we showed that the presence of a time-integral term in IFTs is related to the unseparated measurement and feedback phases.
Since the Szilard-type demons and the 8-state model have the separated measurement and feedback phases, the IFTs for the Szilard engine type demons [\eqqref{SU}] and for the 8-state model [\eqqref{class1eq}] do not contain any time-integral term.
In contrast, since the autonomous bipartite demons have the unseparated measurement and feedback phases, the IFT for the autonomous bipartite demons [\eqqref{class2eq}] contains a time-integral term.

On the basis of the general framework in Sec.~\ref{s:gen-frame}, we clarified the concept of separation as the condition on the choice of the subset of transitions $\Omega$.
The separation condition leads to the absence of time-integral terms, which is clearly shown in \eqqref{sepStates}.
Understanding the separation of the measurement phase and the feedback phase is important to analyze the difference between ideal information processing systems (e.g., the Szilard engine) and demons in the real world such as biochemical networks.

\ack

The authors thank S.-i. Sasa and H. Tasaki for fruitful discussion. 
The authors also thank U. Seifert for helpful comments.
NS, SI, KK were supported by Grant-in-Aid for JSPS Fellows Number 26-7602, 24-8593, 24-8031 respectively, and TS was supported by JSPS KAKENHI Grant Number 25800217, 22340114.
NS and TS are also supported by Platform for Dynamic Approaches to Living System from MEXT, Japan.


\end{document}